# Stabilized narrow beam emission from Broad Area Semiconductor Lasers

JUDITH MEDINA PARDELL, [1] RAMON HERRERO, [1] MURIEL BOTEY, [1] KESTUTIS STALIUNAS[1,2]

[1]*Departament de Física, Universitat Politècnica de Catalunya (UPC), Colom 11, E-08222 Terrassa, Barcelona, Spain*
[2]*Institució Catalana de Recerca i Estudis Avançats (ICREA), Passeig Lluís Companys 23, E-08010, Barcelona, Spain*
*\*Corresponding author: judith.medina@upc.edu*

**Abstract:** We provide a feasible and compact scheme to control and stabilize the spatiotemporal dynamics of BAS lasers. The proposal is based on the ability of non-Hermitian potentials with given local symmetries to manage the flow of light. A local PT-symmetric configuration allows controlling, enhancing and localizing the generated light. We impose a pump modulation, with a central symmetry axis which induces in-phase gain and refractive index modulations due to the Henry factor. Both modulations are, in turn, spatially dephased by an appropriate index profile to yield to a local PT-symmetry within the modified BAS laser. Such local PT-symmetry potential induces an inward mode coupling, accumulating the light generated from the entire active layer at the central symmetry axis, which ensures spatial regularization and temporal stability. By an exhaustive exploration of the modulation parameters, we show a significant improvement of the intensity concentration, stability and brightness of the emitted beam. This approach produces a two-fold benefit: light localization into a narrow beam emission and the control over the spatiotemporal dynamics, improving the laser performance.

Broad Area Semiconductor (BAS) amplifiers and lasers are prevalent and trustworthy light sources, used for many applications ranging from biomedicine [1] to telecommunications [2]. Their main advantage is the compactness and the high conversion efficiency, while their major drawback is the relatively low spatial and temporal quality of the emitted beam [3, 4]. The fundamental phenomenon inducing spatiotemporal instabilities is Modulation Instability (MI) [5] which along with nonlinear modal interaction leads to complex spatiotemporal dynamics and filamentation, limiting possible applications [6, 7]. Possible stabilizations of BAS sources propose introducing spatial [8.9] or spatiotemporal modulations [10], or spatial non-Hermitian potentials, with in-phase spatial modulations of refractive index and pump, leading to a substantial improvement of beam quality [11]. In addition, vertical-external-cavity surface-emitting lasers with external flat mirrors can be stabilized by applying a periodic spatiotemporal modulation of the pump current [12]. More recently, attention was paid on a specific kind of non-Hermitian systems, holding PT-symmetry [13, 14]. In such materials, the complex refractive index fulfills: $n(x) = n^*(-x)$ i.e., the real part representing the refractive index is symmetric while the imaginary part representing gain-loss is antisymmetric in space. In periodic PT-symmetric media index and gain-loss modulations are dephased by a quarter of the wavenumber of the modulation. Such systems hold maximum asymmetric mode coupling at the PT-symmetry breaking point, when the gain-loss and refractive index modulation amplitudes are balanced [15, 16]. In optics, global or local PT-symmetry leads to unconventional beam dynamics arising precisely from the asymmetric mode coupling, such as the unidirectional light transport following arbitrary vector fields [17, 18]. Therefore, we here propose that local PT-symmetry could also be applied to regularize the spatiotemporal dynamics of BAS lasers, while improving stability through unidirectional mode coupling.

In this letter, we propose to apply a local PT-symmetric potential to control the complex spatiotemporal dynamics of BAS lasers, see Fig. 1a, by localizing the light generated in the entire active layer into a narrow central beam, see Fig.1b. The proposal is twofold: to regularize and control the emission of modulated BAS leading to a narrow beam emission, expected to be useful for a larger variety of practical applications. The proposed model is feasible and could be implemented by a simple patterning of the metal layer [19] for the pump modulation and using well-known nanofabrication techniques to structure the semiconductor substrate as in [20, 21], since, as it will be shown, the required index periodic modulation is on the order of $\Delta n \approx 10^{-3}$ matching the spatial modulation amplitudes of the proposed modulations.

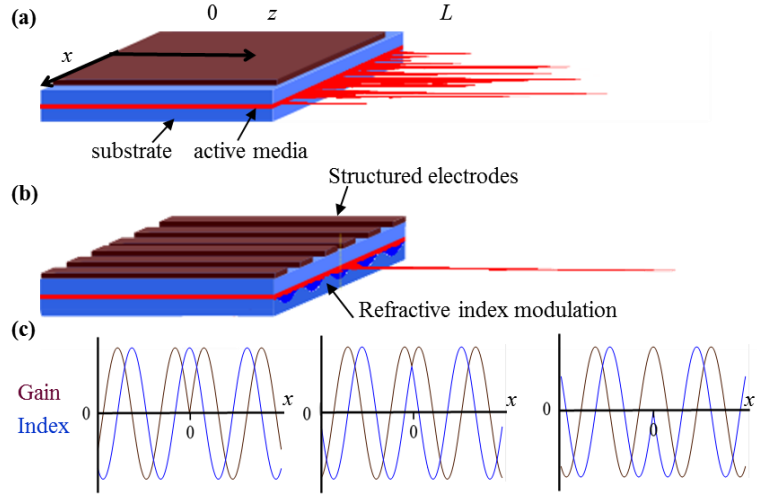

Fig.1. BAS emission, schematic illustration: (a) complex irregular spatial pattern emitted from a conventional BAS source, (b) bright and narrow beam from a modified BAS with local PT-symmetry. (c) Scheme of the transverse spatial distribution of the local PT-symmetric potential, with balanced gain (in brown) and effective refractive index (in blue) modulations, for different configurations.

BAS semiconductor sources are generally described by stationary models including field and carriers [22] or, alternatively, by mean field models including temporal evolution [23]. A different approximation for BAS amplifiers is given by the optical field propagation along the semiconductor, while carrier and field dynamics is neglected. Previous studies on local PT-symmetry were similarly performed considering strongly simplified stationary models and only considering forward field propagation in a paraxial approximation [17].

These simplifications become inadequate for the spatial regularization and temporal stabilization of BAS lasers that requires a spatiotemporal calculation of field and carriers. Thus, we develop a new integration scheme taking advantage of the different time scales of the cavity round trip time (in the order of ps) and carrier's relaxation time (in the order of ns). Every integration step combines the field propagation in one cavity round trip assuming constant carriers and the temporal integration of carriers, considering an already stabilized field. The following BAS source representation obtained from well-known models [22, 23], allows integrating along the cavity the field amplitude envelope, $A$, —composed by the forward, $A^+$, and backward, $A^-$, fields—, and to integrate carriers, $N$, in time:

$$\pm\frac{\partial A^{\pm}}{\partial z}=\frac{i}{2k_{0}n}\frac{\partial^{2}A^{\pm}}{\partial x^{2}}+s[(1-ih)N-(1+\alpha)]A^{\pm}+i\Delta n(x,z)k_{0}A^{\pm} \quad (1)$$

$$\frac{\partial N}{\partial t}=\gamma(-N-(N-1)|A|^{2}+p_{0}+\Delta p(x,z)+D\nabla^{2}N)$$

where $|A|^{2}=|A^{+}|^{2}+|A^{-}|^{2}$

and where $k_0$ is the wavevector, n the refractive index, $s$ a parameter inversely proportional to the light matter interaction length, $h$ the Henry factor (linewidth enhancement factor of the semiconductor), α corresponds to losses, $\gamma$ is the carriers relaxation rate, $p_0$ the pump, and $D$ is the carrier diffusion. The polarization of the semiconductor is adiabatically neglected (as typically considered for class B lasers). Finally, $\Delta p(x,z)$ and $\Delta n(x,z)$ represent the pump and index modulations, respectively. We assume a symmetric harmonic transverse modulation of the pump in the form: $\Delta p(x,z) = m_1 \sin(q_x|x|+\Phi)$ where $q_x = 2\pi/d_\perp$, being $d_\perp$ the transverse period, and $m_1$ the amplitude of the pump modulation ($z$ is the propagation direction). Such pump modulation induces, in turn, an in-phase refractive index modulation trough the Henry factor. Therefore, the introduced harmonic refractive index modulation: $\Delta n(x,z) = m_2 \cos(q_x|x|+\Phi) + m_3 \sin(q_x|x|+\Phi)$ is intended to render the refractive index in quadrature with the refractive modulation (first term) and to compensate the induced gain modulation (second term). The potential has a symmetry axis at $x = 0$, which divides the system in two half-spaces, both holding a global PT-symmetry. Such potential is expected to couple light asymmetrically in the transverse direction, promoting the inward coupling, enhancing and localizing light at axis. The modulated BAS system depends on four parameters, namely the three modulations amplitudes ($m_1$, $m_2$, $m_3$) and the phase, $\Phi$ that controls the character of the center — in Fig. 1.c the modulation profiles correspond to: $\Phi = 0$ (left), $\Phi = \pi/4$ (center) and $\Phi = \pi/2$ (right) —.

We preliminarily analyze a BAS amplifier with real parameters as a proof of concept of the regularization scheme based on local PT-symmetry, before considering the management of radiation from laser sources.

The local PT-symmetric BAS amplifier is modelled by the system of equations (1), only assuming a forward propagating field, i.e. $A^- = 0$. The spatiotemporal integration of the unmodulated BAS exhibits both inhomogeneous spatial and unstable temporal behaviors (Figs.2.a and 2.b). Introducing the above-mentioned pump and refractive index periodic modulations with PT-symmetric profiles, the BAS radiation emission may be regularized to enhance and spatially localize light at the symmetry center, resulting into a bright and narrow BAS amplifier emission (Fig.2.c). Moreover, in addition to spatial regularization, the field becomes temporally stable, see Fig.2.d. Therefore, local PT-symmetry has a direct impact on both spatial and temporal stability. To assess the performance of the proposed modulated BAS amplifier, we evaluate the axial concentration factor, as a figure of merit of light localization, defined as the central intensity, $I_0 = I(x = 0)$, over the averaged intensity: $c = I_0 /< I(x)>$. We explore the parameter space of the index modulations amplitudes $m_2$ and $m_3$, for a fixed value of the pump modulation amplitude, $m_1$, and a fixed phase, $\Phi = 0$. We expect that for particular parameters, the inward coupling localizes light around $x = 0$, while the central index maximum contributes with a guiding effect. The results are summarized on Fig. 2.e. Depending on the sign of the product $m_1 \cdot m_2$, we may expect regions of inward or outward mode coupling. The inward coupling ($m_1 \cdot m_2 > 0$) leads to an accumulation of the field around the center, which is maximized for a particular value of the counterbalancing amplitude: $m_3$ (being also $m_1 \cdot m_3 < 0$). Note that the maximum energy localization is expected for balanced gain and refraction index modulations in a local PT-symmetric potential [17]. Indeed, for $m_1 > 0$, we observe the maximum confinement for positive $m_2$ and negative $m_3$ values (Fig. 2.e). For $m_2 < 0$ the concentration factor is below 1 indicating that the field at the center is lower than the average, consistent with the outward coupling. Finally, Fig.2.f depicts the transverse beam intensity profile for three particular situations, corresponding to the same values for $m_1$ and $m_3$ but for positive, zero and negative values of $m_2$. We may observe that when the modulations induce an inward coupling light is localized and strongly enhanced at the center. On the contrary, when index and gain modulations are in phase ($m_2 = 0$) the coupling is symmetric, and the field is homogeneously distributed, and when coupling is outwards ($m_2 < 0$) intensity at the center is reduced.

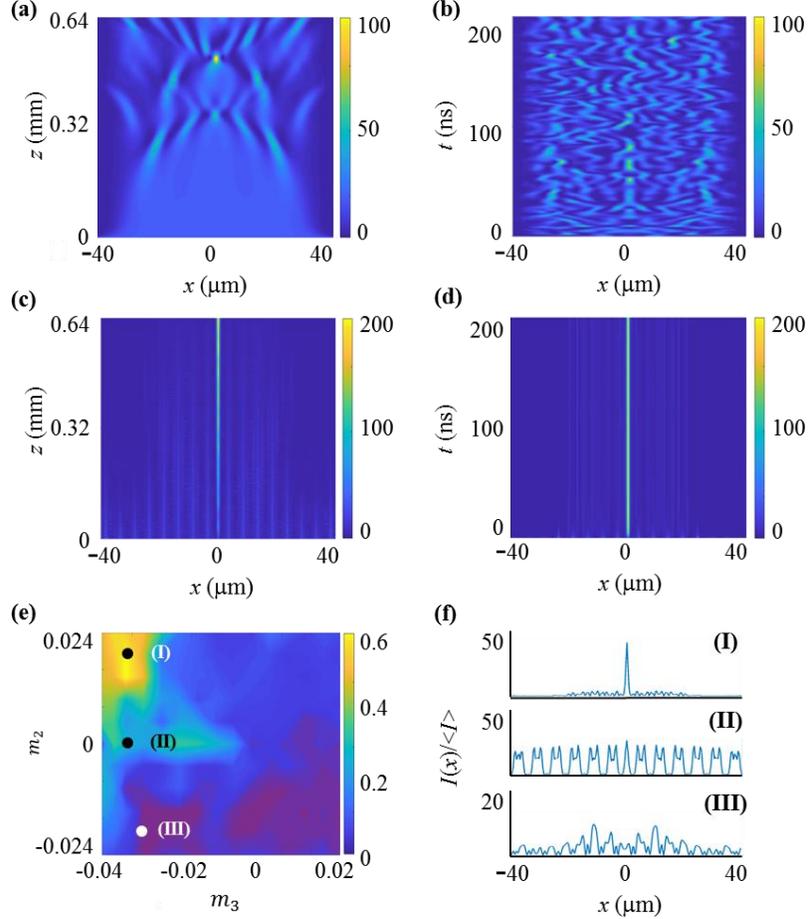

Fig.2. Unmodulated BAS amplifier: (a) spatial intensity distribution and (b) temporal evolution of the intensity for $p_0 = 1.23$ (arbitrary units). Local PT-symmetric BAS amplifier: (c) spatial field localization, and (d) stable, bright and narrow emission for $m_1 = 0.5$, $m_2 = 0.021$ and $m_3 = -0.029$, $\Phi = 0$ and $p_0 = 1.23$. (e) Axial concentration factor map in ($m_2, m_3$) space for $m_1 = 0.5$. (f) Transverse cuts of the intensity distributions, at $z = L$, normalized to the mean intensity, for three representative points of the map, corresponding to: (I) PT-symmetric distribution leading to inwards mode coupling, $m_2 = 0.021$ and $m_3 = -0.033$; (II), symmetric mode coupling, in phase gain and index modulation $m_2 = 0$ and $m_3 = -0.033$; (III) local PT-symmetric distribution leading to outwards mode coupling $m_2 = -0.021$ and $m_3 = -0.029$. Other integration parameters: $L = 640 \mu m$, $\alpha = 0.1$, $\mu m^{-1}$, $h = 2.0$, $s = 0.01 \mu m^{-1}$, $q_x = 1.2566 \mu m^{-1}$, $k_0 = 2\pi \mu m^{-1}$, $D = 0.03$ cm$^2$/s, $\gamma = 0.01$.

The BAS laser is also described by the system of equations (1) where the field components, $A^+$ and $A^-$, are related by the corresponding boundary conditions $A^+(x, z = L, t) = r_L A^-(x, z = L, t)$ and $A^-(x, z = 0, t) = r_0 A^+(x, z = 0, t)$. Where $L$ is the semiconductor length and $r_{0/L}$ are the corresponding reflection coefficients of the cavity mirrors at $z = 0/L$, respectively, see Fig.3.a.

The effect of the phase, $\Phi$, is more critical for the local PT-symmetric BAS laser, due to the feedback introduced by the cavity. The optimum situation for field concentration is $\Phi = \pi/4$, since gain-maxima are located closer to the center, while an index relative maximum is still present at $x = 0$ preserving an index guiding effect. For $\Phi = 0$ and $\Phi = \pi/2$, larger modulations of the complex refractive index (larger values of $m_2$ and $m_3$) are needed to obtain an effective inward coupling. Analogously to the preliminary BAS amplifier study, the radiation from BAS lasers holding PT-symmetry may be spatially regularized into a bright beam, see Fig.3.a. and Fig.3.b. We perform a comprehensive exploration of the axial concentration; the results are mapped in Fig.3.c. We observe regions of maximal localization for $m_2 > 0$ and $m_3 < 0$, when the coupling is inwards, achieving a significant concentration. We note that the required $m_2$ and $m_3$ amplitudes are even smaller than the required for the spatiotemporal stabilization of a corresponding BAS amplifier. This is attributed to the long effective cavity length $L/(1-r_0)(1-r_L)$, on the order of $10^2 \cdot L$ for the considered reflectivities, see Fig.3.d. Finally, we note that for $\Phi = \pi/4$ the area of maximal axial concentration, mapped on Fig.3.c, is slightly shifted as compared to Fig.2.e.

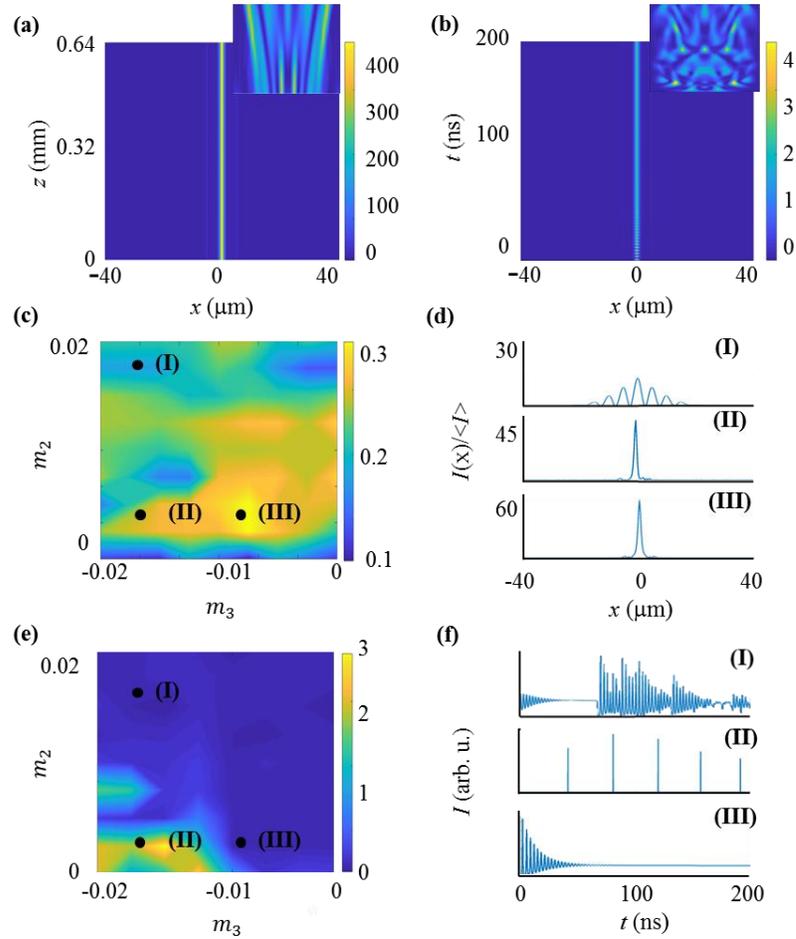

Fig.3. (a) Localized spatial intensity distribution (in W/cm$^2$) within the modulated BAS laser and (b) temporal evolution of the narrow beam emission from the modified BAS with $m_1 = 0.5$,

$m_2 = 0.0048$ and $m_3 = -0.0143$ with $p_0 = 1.23$, $\Phi = \pi/4$, $r_0 = 0.99$ and $r_L = 0.9, \gamma = 0.005$. The insets depict the inhomogeneous laser spatial intensity pattern and unstable laser emission of the unmodulated case, respectively. (c) Axial concentration factor mapped for $m_1 = 0.5$ in the $(m_2, m_3)$ parameter space. (d) Intensity distribution transverse cuts, at $z = L$, normalized to the mean intensity, for three representative points corresponding to: (I) $m_2 = 0.0143$, $m_3 = -0.0143$, (II) $m_2 = 0.0048$, $m_3 = -0.0143$, (III) $m_2 = 0.0048$, $m_3 = -0.0072$. (e) Temporal stability map of the PT-symmetric BAS laser: amplitude of the temporal intensity oscillations. (f) Temporal evolution of the emission for the previous three representative points, (I) and (III) achieve a temporally stable regime while (II) corresponding to a pulsed regime. All other integration parameters are the same as in Fig.2.

In addition, we assess the temporal stability of the emitted narrow beam by mapping the temporal oscillations of the peak intensity. Inspecting Fig.3.e, we observe that not all the spatial concentration region on Fig.3.c is temporally stable, yet there is a wide range of parameters for which we simultaneously achieve full regularization of the emission, spatial field localization and temporal stability. Temporal instabilities arising from pulsed regimes appear for a restricted island of parameters, see Fig.3.f. The laser is temporally stable at the maximum concentration area corresponding to the set of parameters (III). Typical relaxation oscillations at the order of ns, the carriers' characteristic time, are visible for this case at pump values just above threshold. For other cases, the laser becomes unstable, with large amplitude oscillation after some μs. Simulations also show that the temporal stabilization is mainly controlled by the balance between pump and refractive index modulations, namely the value of $m_2$, while the spatial shift between potentials, $m_3$, determines spatial instability. Besides, it can be easily checked that a narrower laser would lead to smaller output intensity. We have performed a series of calculations to assess the laser output dependence on the transverse size (for a constant transverse modulation). The results are summarized in Fig.4.

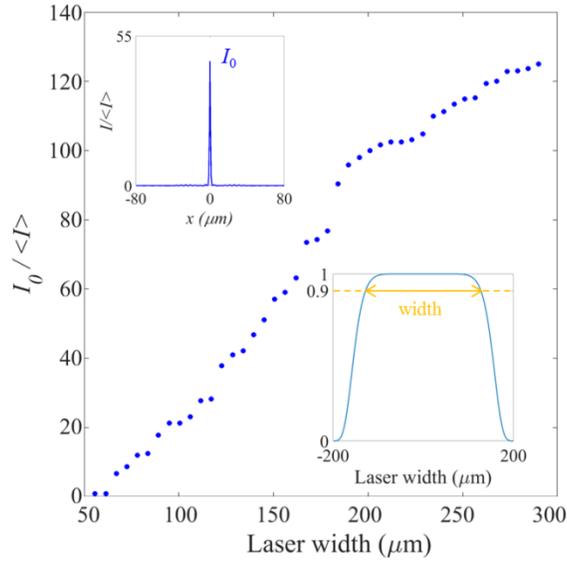

Fig.4. Dependence of the laser intensity concentration on the laser width. Upper inset: example of normalized spatial beam profile, for a laser width of 160 μm. Lower inset: the laser width is determined by the transverse pump profile at 0.9 the maximum pump. Integration parameters: $m_1 = 0.5$, $m_2 = 0.021$, $m_3 = -0.033$, $L = 640$μm, $\alpha = 0.1$, μm$^{-1}$, $h = 2.0$, $s = 0.01$ μm$^{-1}$, $q_x = 1.2566$ μm$^{-1}$, $k_0 = 2\pi$ μm$^{-1}$, $D = 0.03$ cm$^2$/s, $\gamma = 0.005$ and $r_L = 0.04$.

We finally analyze the BAS laser general performance in different working conditions -as a function of the input homogeneous pump, $p_0$, and pump modulation amplitude, $m_1$, keeping the index modulation amplitudes, $m_2$ and $m_3$, optimized and proportional to $m_1$. As might be expected, increasing $m_1$, the lasing threshold of the central area decreases leading to two clear different lasing regimes, see Fig.5a. For small pump powers, lasing is mainly restricted to the central area with a high intensity concentration factor, see Fig.5b. For sufficiently high modulations a bright and narrow beam is generated even below the homogenous laser threshold ($p_0 = 1.2$). Above threshold, amplification occurs in all the active material, and localization persists. The mean generated intensity increases with the pump, almost independent of the potential, either for the unmodulated or modulated BAS lasers, see the dotted curves in Fig.5.c. We note that for small modulation amplitudes of the local PT-symmetric potentials the peak intensity $I_0$ grows faster, increasing pump while having less concentration. This is attributed to the existence of several transverse modes for the peak profile, see Fig.5.d. Moreover, the lasing threshold curve in Fig.5.a shows clear different slopes in the $p_0$-$m_1$ plane for both peak transverse profiles. The transition between these profiles does neither show hysteresis nor bistability (Fig.5.d). We note that these calculations while performed for a set of parameters are restricted to realistic values.

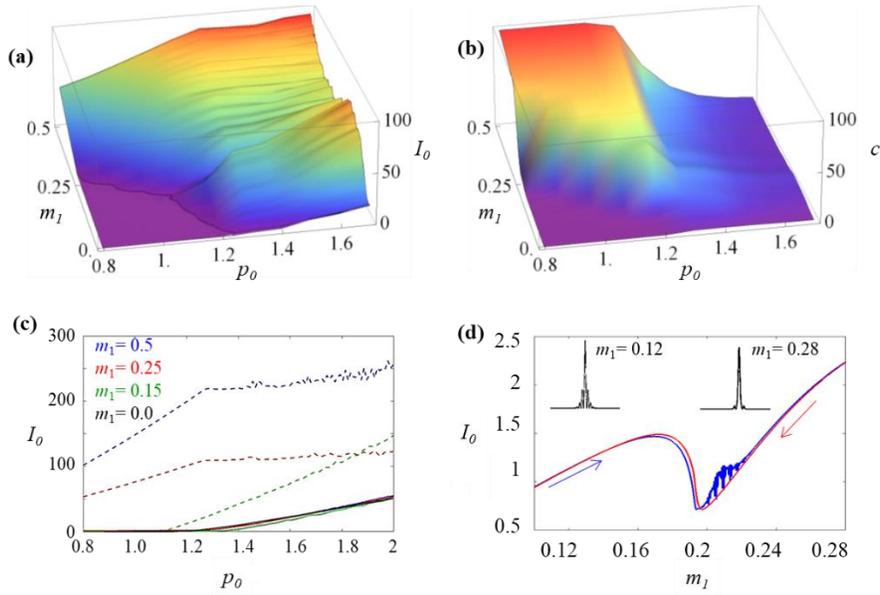

Fig.5. (a) Intensity at $x = 0$, $I_0$, as a function of pump, $p_0$, and the pump modulation amplitude $m_1$, assuming the optimized corresponding values for the index modulations amplitudes $m_2$ and $m_3$. The solid line indicates the pump threshold for lasing. (b) Intensity concentration, $c = I_0/\langle I(x)\rangle$. (c) Central intensity $I_0$ (dashed curves) and mean intensity $\langle I(x)\rangle$ (solid curves) for different values of the pump $p_0$, as a function of $m_1$. The unmodulated case (black curve) is depicted for comparison. (d) Central intensity for $p_0 = 1.23$ showing a transition around $m_1 = 0.2$ with different transverse mode localization profiles, when integration is performed increasing/decreasing the $m_1$ value. All other parameters are the same as in Fig.3.a. except $r_L = 0.04$.

To conclude, we propose local PT-symmetric potentials to tailor and control the complex spatiotemporal dynamics of BAS lasers. We propose to introduce modulations in the pump and refractive index with a central symmetry axis to induce local PT-symmetry. We propose a simple symmetric harmonic modulation of the pump, and dephased index modulation intended to generate a symmetric and local PT-symmetry and to compensate the index modulation induced by the pump modulation trough the Henry factor. The field regularization mechanism relies on the inward coupling of the light generated in the active layer, which directs and concentrates light towards the center. The maximum localization regimes are first analyzed for a BAS amplifier. A significant intensity field enhancement and concentration of the emitted beam are found as compared to the unmodulated amplifier, coinciding with inward PT-symmetric coupling. The results are further numerically extended to the more engaged case of the BAS laser, where both forward and backward field components are considered. In this case, the character of the symmetry center (controlled by a general phase, $\Phi$) is more significant. We show that the proposed scheme is able to simultaneously regularize the radiation in a bright and narrow beam with temporal stabilizations is achieved, in a twofold benefit. The system is studied under different working conditions observing a substantial filed regularization, especially for pumps below and close to the unmodulated laser threshold. The proposed scheme renders BAS lasers into bright and stable sources. Moreover, the proposed scheme is general and can be applicable to other BAS lasers to improve their performance.